\documentclass[10pt,aps,apl,twocolumn,superscriptaddress,floatfix,showpacs,longbibliography, titlepage]{revtex4-1}

\usepackage{graphicx}
\usepackage{braket}
\usepackage[dvipsnames]{xcolor}
\usepackage{multirow}
\usepackage{makecell}
\usepackage{amsmath}
\usepackage{amssymb}
\usepackage{dcolumn}
\usepackage{bm}
\usepackage{hyperref}

\hypersetup{hidelinks}

\date{\today}

\begin{document}
	
	\title{Neural network agent playing spin Hamiltonian games on a quantum computer}
	
	\author{Oleg M. Sotnikov}
	\email[E-mail for correspondence: ]{oleg.sotn@gmail.com}
	\author{Vladimir V. Mazurenko} 
	
	\affiliation{Theoretical Physics and Applied Mathematics Department, Ural Federal University, 620002 Ekaterinburg, Russia}
	
	
	\begin{abstract}
		Quantum computing is expected to provide new promising approaches for solving the most challenging problems in material science, communication, search, machine learning and other domains. However, due to the decoherence and gate imperfection errors modern quantum computer systems are characterized by a very complex, dynamical, uncertain and fluctuating computational environment. We develop an autonomous agent effectively interacting with such an environment to solve magnetism problems. By using the reinforcement learning the agent is trained to find the best-possible approximation of a spin Hamiltonian ground state from self-play on quantum devices. We show that the agent can learn the entanglement to imitate the ground state of the quantum spin dimer. The experiments were conducted on quantum computers provided by IBM. To compensate the decoherence we use local spin correction procedure derived from a general sum rule for spin-spin correlation functions of a quantum system with even number of antiferromagnetically-coupled spins in the ground state. Our study paves a way to create a new family of the neural network eigensolvers for quantum computers.
	\end{abstract}

\maketitle

Reinforcement machine learning techniques were initially developed for creating autonomous intelligent robotic systems \cite{thesis}. 
Currently, they are being successfully applied in completely different decision making domains such as games \cite{game,Go1,Go2}, traffic control systems \cite{traffic}, computer resources management \cite{management}, news recommendation systems \cite{news}, optimization of chemical reactions \cite{chemical} and others. Within a reinforcement learning approach an agent taking some actions interacts with environment, receives feedback, estimates rewards and corrects its actions to increase a future reward. This idea is very attractive, since in such a formulation the agent is fully autonomous and can develop different strategies to gain more. However, a practical realization of a reinforcement learning technique is problem specific and requires additional innovations providing the stability and convergence of numerical schemes.

Reinforcement learning techniques have been actively developed and implemented in such a new field of research as quantum computing. It includes quantum-error-correction systems in complex quantum devices \cite{petru,qerr1,qerr2,qerr3}, design and implementation of quantum communication technologies \cite{arxiv1}, quantum gate control \cite{arxiv2,arxiv21, arxiv22,arxiv23}, quantum gate design \cite{arxiv3}, quantum algorithms for reducing computational error \cite{arxiv4}. It is important to note that only a few of these algorithms were tested on real quantum devices.

Motivated by recent results of Google DeepMind team \cite{game} obtained for classic Atari games in this work we develop and practically realize a reinforcement learning scheme for approximating the ground states of spin Hamiltonians on quantum computers. In this field of quantum computing there are two approaches widely used to simulate magnetic systems. The first one is an adiabatic simulation method \cite{evolu1, evolu2, evolu3, Pogosov} that is based on the discretization (Trotterization) of the time evolution operator. Another one is a variational quantum eigensolver proposed in Refs.~\onlinecite{VQE,Troyer, Malley} that uses the Ritz's variational principle to prepare approximations of the ground state of a magnetic model. 
Being different by the construction both approaches assume to use some fixed sequences of quantum gates. 

Practical applications of these methods for the simplest quantum spin Hamiltonian on real home-made and public quantum devices \cite{Pogosov, IBMQE, IBM1, evolu2} have revealed problems that are mainly related to the decoherence and gate errors. For instance, the experiments aimed to spin Hamiltonian dynamics \cite{Pogosov} have demonstrated that such errors become more and more significant as the length of the quantum program increases. As a result, a few Trotter steps lead to a considerable error in the experimental data in comparison with exact results. On the other hand, the variational procedure \cite{VQE} requires a calibration of gradient-descent method parameters to probe the energy landscape in the vicinity of the state defined with current set of parameters. These additional measurements are also source for errors.    
 
In our study we follow a distinct logic and consider a spin Hamiltonian problem as a game with the following rules. Starting with a random quantum state a player performs several quantum actions and measurements to get the best score that means the lowest energy and, as a result, the best approximation of the spin Hamiltonian ground state. To play this game we develop a multi-neural-network agent that  determines a sequence of quantum gates for a short quantum circuit. In contrast to previous approaches we do not use a fixed sequence of quantum gates, and at each iteration the agent chooses a new gate for quantum circuit depending on the current state of a quantum device on the basis of the calculated correlation functions. During the training process the agent writes short quantum programs and runs them on a simulator with noise. Here we apply the best latest expertise in the field of the reinforcement learning \cite{game}. For instance, we use the experience replay mechanism repeatedly presenting the past experiences to its learning algorithm and an iterative update procedure adjusting the action-values (Q) towards target values. They provide the stability of the whole scheme running on the noisy quantum systems that are simulator and real device. Having trained the agent on the quantum simulator by using the developed reinforcement learning technique we demonstrate its performance on real IBM Quantum Experience  devices.

\section*{1-qubit problem}
We start with description of a single-spin Hamiltonian problem to explain details of our approach.  The Hamiltonian is given by $\hat {\rm H} = {\bf B} \hat {\bf S}$, where ${\bf B}$ is the external magnetic field and $\hat {\bf S}$ is the spin-$\frac{1}{2}$ operator.
The components of the external magnetic field were chosen to be ${\bf B}$ =(1,1,1). The solution of the problem can be obtained with universal $U_3 (\theta,\phi,0)$ gate that acts on a qubit in the initial state, $\ket{0}$. Here $\theta \approx 2.186276$ rad  and $\phi = -\frac{3\pi}{4}$. It gives the ground state $\ket{\Psi_0}= U_{3} \ket{0}$. The definitions of the gates we use are given in the Supplementary Material \cite{supplementary}.    
\begin{figure}[!t]
	\includegraphics[width=\columnwidth]{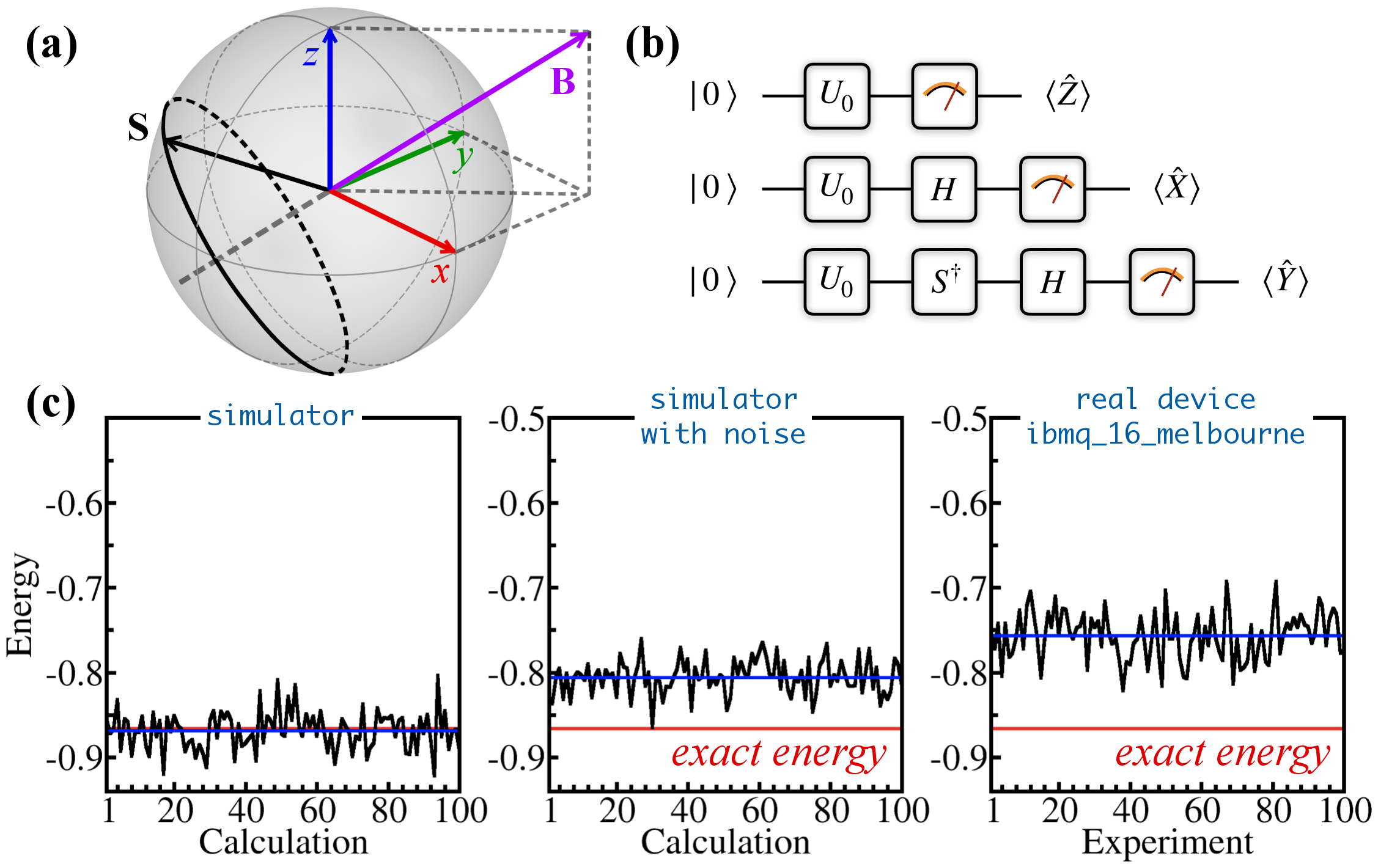}
	\caption{(a) Schematic visualization of a quantum spin in the magnetic field $\mathbf{B}$ = (1,1,1). (b) Quantum circuits used to calculate the correlation functions and energy of the system. (c) The energies obtained from 100 independent calculations on the quantum simulator (left and central panels) and 100 experiments on the real quantum device (right panel). The number of measurement shots was equal to 1024. Blue lines correspond to the average energies. The energy of the exact solution depicted with red line is equal to $E_0$= -0.866.}
	\label{ground-noise}
\end{figure}

For the considered single spin problem the energy is given by 
\begin{eqnarray}
\label{energy}
E = \frac{1}{2}(B^x \braket{\hat X} + B^y \braket{\hat Y} + B^z \braket{\hat Z}),
\end{eqnarray} 
where $\braket{\hat Z}$, $\braket{\hat X}$ and $\braket{\hat Y}$ are the correlation functions calculated by using the probabilities of the basis states, which is a standard output of a quantum computer. It is important to note that these correlators estimated with 
\begin{figure}[!h]
	\centering
	\includegraphics[width=0.6\columnwidth]{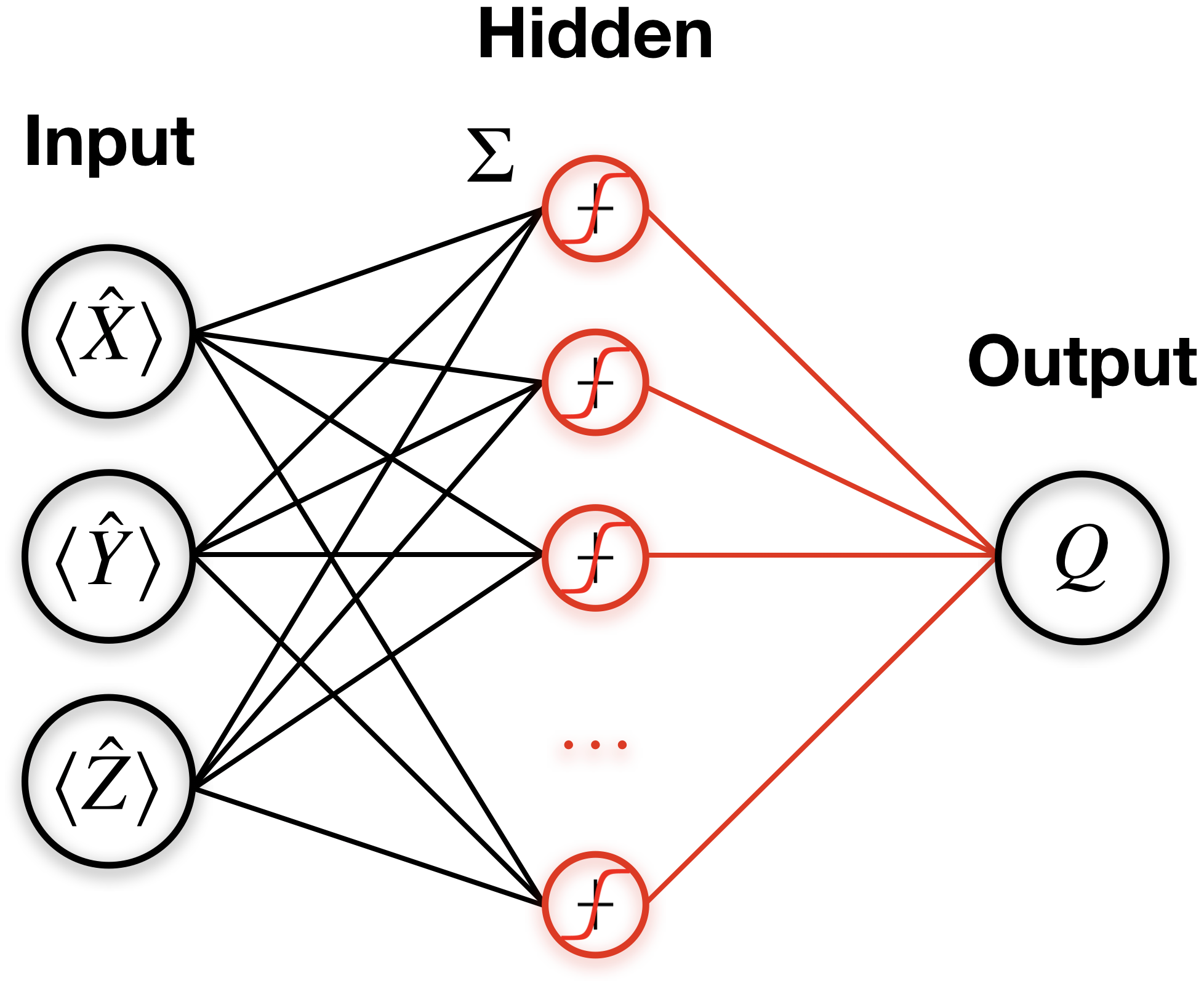}
	\caption{Schematic representation of the neural network we use to define the action-value for each gate.}
	\label{energy_and_reward}
\end{figure}
real quantum computer are subjected by decoherence and gate imperfection errors. The simulations of these errors is an active field of research \cite{noise1,noise2,noise3}. To take these effects into account we used a noise generator as implemented in Qiskit Aer module \cite{noise_Aer}. It includes gate and readout errors imitating the real quantum device noise approximated as a relaxation process of qubits involved in the experiment. Despite the noise model of the local simulator is simple and takes into account only local errors occurring on each gate, we will show that inclusion of device noise even on this level significantly improves the agreement between results of simulations and experiments obtained with real device. We used a basic device noise model with parameters reported in official Qiskit tutorials for IBM Q 16 Melbourne device~\cite{qiskit_tutorials}. 

With circuits presented in Fig.~\ref{ground-noise} we have performed 100 independent calculations on simple quantum simulator, 100 calculations on the simulator with noise and 100 experiments on the real quantum device. In all the cases we observe fluctuations of the measurements results. For instance, for the  simple simulator the energy fluctuates around exact value. The account of decoherence and gate imperfections with the noise model leads to a higher average energy that is about -0.8. Real experimental results obtained with IBM Q 16 Melbourne device are characterized by very strong fluctuations of the energy around -0.75.  

These experiments clearly demonstrate that even for the simplest problem one deals with complex, dynamical, uncertain and fluctuating quantum computing environment. It motivates to develop an autonomous agent effectively interacting with such an environment.

\subsection*{Neural network agent}
The agent we develop is multi-network one in according with a one-action-one-network concept proposed in Ref.~\onlinecite{thesis}. There is a separate network for each action, but the structures of all the networks are the same (Fig.~\ref{energy_and_reward}). They contain input, one hidden and output layers. The network takes spin correlation functions obtained with quantum computer or simulator as an input. The number of the input neurons depends on the problem we solve.  For instance, in the case of the one-qubit problem (single spin in an external magnetic field) there are three correlation functions that form a state vector ${\bf s} = \{ \braket{\hat X}, \braket{\hat Y},\braket{\hat Z} \}$ characterizing the quantum system in question. In turn, there will be 6 single-spin and 9 spin-spin correlation functions (15 in total) in the case of the dimer spin Hamiltonian problem.

The number of the neurons in the hidden layer also depends on the size of the problem. In the one- and two-qubit cases we consider the number of the hidden neurons is equal to 32 and 64, respectively. The output layer contains only one neuron that represents the predicted action-value Q for the particular quantum gate (action). Having compared the calculated Q-values among all the networks the agent chooses the gate for which action-value function has the maximal value.
\begin{figure}[!h]
	\includegraphics[width=\columnwidth]{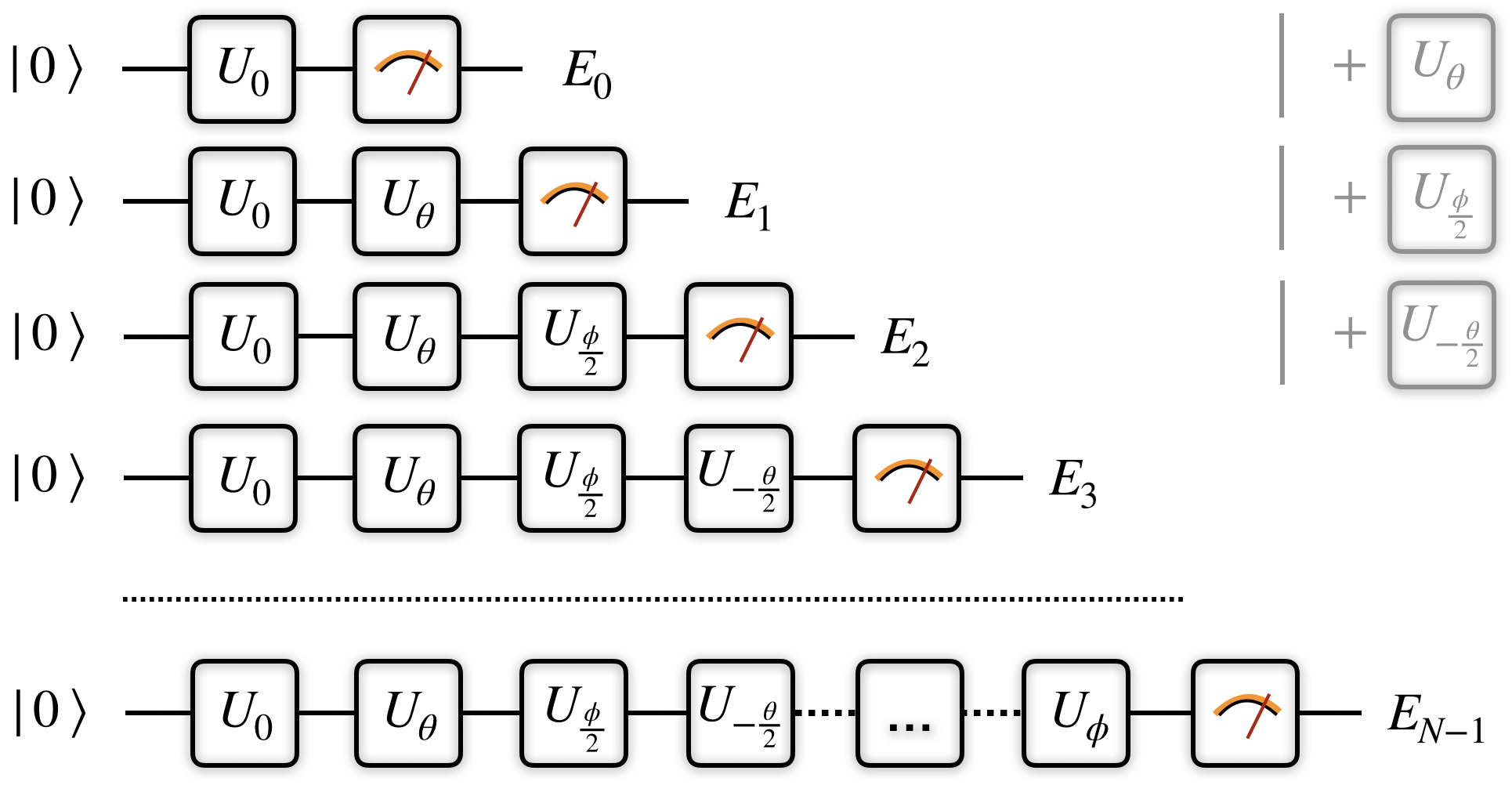}
	\caption{Step-by-step construction of a quantum circuit of N gates. $U_0 = U_3^{rand}$ denotes preparation of a random initial state. Each line corresponds to adding new gate chosen by the agent from the action list described in the text. $ E_t $ represents the energy estimated at the $t$th step of the circuit formation.}
	\label{program_growth}
\end{figure}
In this way one defines an optimal $Q$-function satisfying \cite{thesis}
\begin{eqnarray}
Q({\bf s},a) = r + \gamma {\rm max} \{ Q({\bf s'},k), k \in actions \}.
\label{action-value}
\end{eqnarray}
According to this expression the utility of an action $a$ in response to a state ${\bf s}$ equals to immediate reward $r$ plus the best utility that can be obtained from the next state ${\bf s'}$ discounted by factor $\gamma$. During reinforcement learning, the difference between the two sides of Eq.~\eqref{action-value} is to be minimized using a back-propagation algorithm \cite{supplementary}.

\subsection*{Quantum computer programming}
The developed agent is aimed to define a quantum circuit according to the predicted action values Q. In the case of the one-qubit problem we use the following set of gates: identity gate, $ U_\theta $, $ U_\phi $, $ U_{-\theta} $, $ U_{-\phi} $, $ U_{\theta/2} $, $ U_{\phi/2} $, $ U_{-\theta/2} $ and $ U_{-\phi/2} $ that are defined with the universal rotation gate $U_{3}$ as described in Ref.~\onlinecite{supplementary}.   

The process of the quantum circuit construction can be demonstrated by the example of Fig.~\ref{program_growth}. We start with an universal $U_3$ gate with random angles. Having added a new gate to the quantum circuit the measurements of the correlation functions are performed. It means that one can trace the energy evolution as the length of the quantum circuit increases.

\begin{figure}[!h]
	\includegraphics[width=\columnwidth]{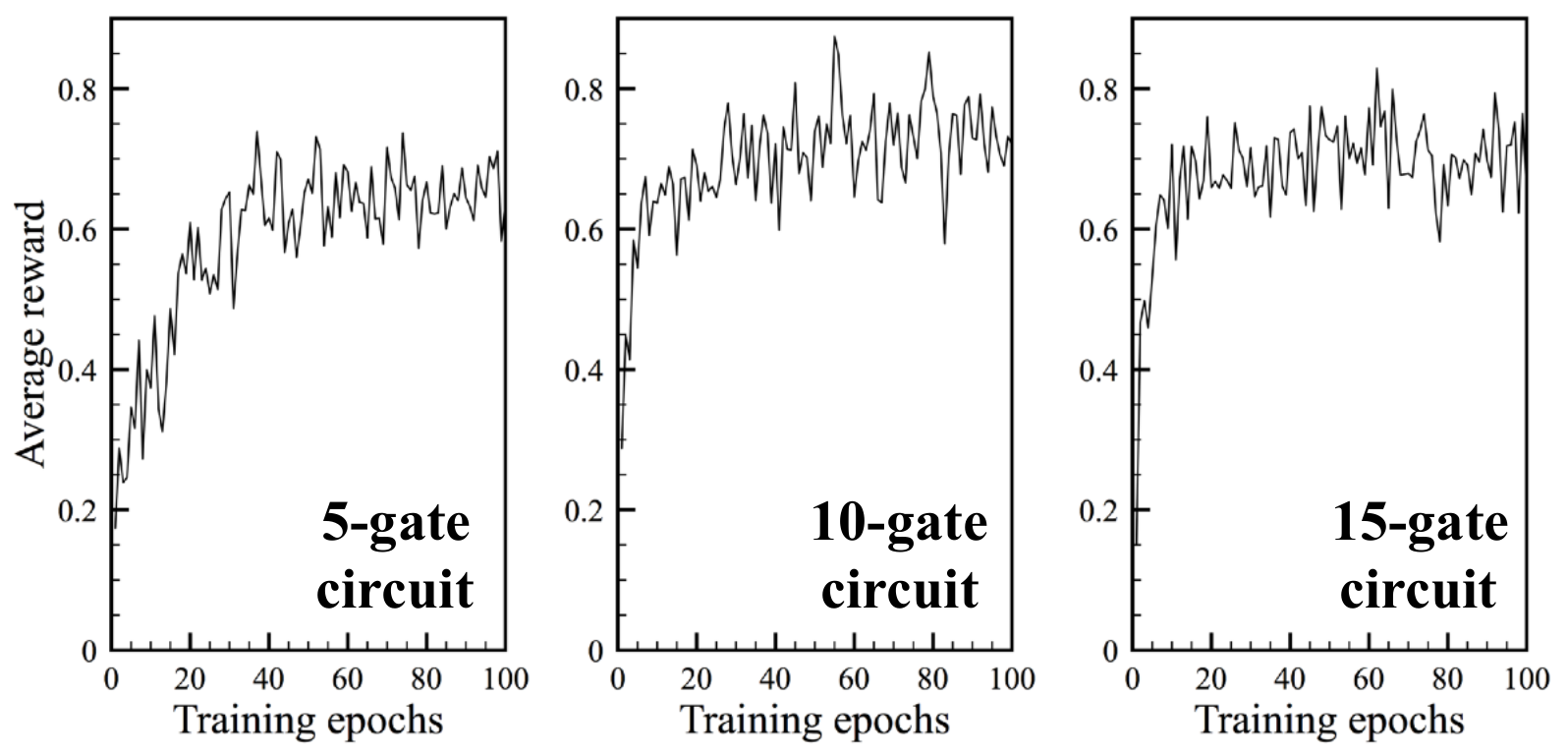}
	\caption{Training results for the simplest problem, single spin in the magnetic field, ${\bf B}$=(1,\,1,\,1). Average reward achieved per episode after the agent is run with $\epsilon$-greedy policy ($\epsilon$=0.05). }
	\label{training_1qubit}
	\vspace{0.5cm}
\end{figure}

\begin{figure*}
	\includegraphics[width=\linewidth]{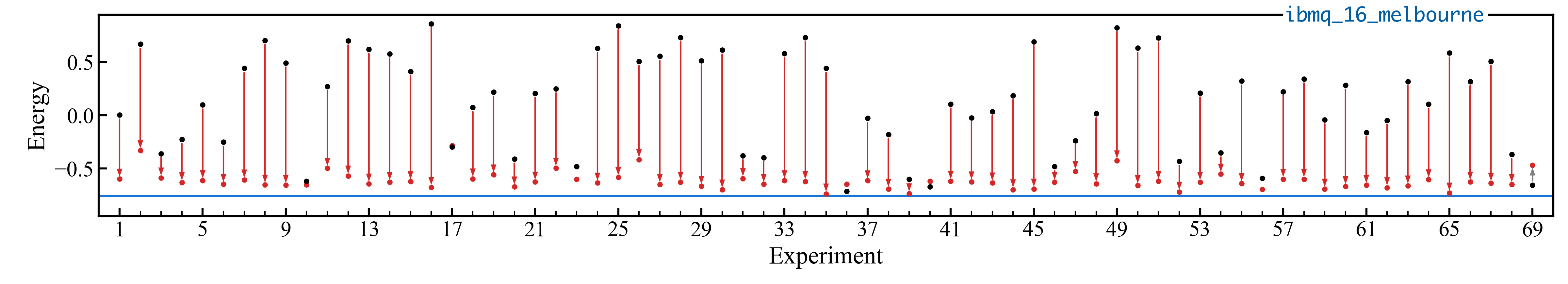}
	\caption{Performance demonstrated by the trained agent in experiments on the IBM Q 16 Melbourne. Black and red circles denote initial and final energies for each 10-gate circuit constructed by the agent. Red and grey arrows indicate lowering and rising of energy with the corresponding circuits. Blue line shows the average energy obtained with exact solution quantum circuits presented in Fig.~\ref{ground-noise} on the real device. The number of measurement shots was equal to 1024.}
	\label{efficiency}
\end{figure*}

\subsection*{Agent's training}
The agent was trained with the IBM QE simulator including the noise model.  
Each training iteration contains three main steps. (i) The agent takes some action following $\epsilon$-greedy policy.
Having added a new gate to the quantum circuit the agent estimates the reward from the observation, $r = E_{t} - E_{t+1}$. (ii) The sequence (state, action, reward, new state) explored by the agent is stored in the replay memory. (iii) A sequence randomly chosen from the memory is used to optimize the weights of the neural network. A complete description of technical details on the agent's training is presented in Ref.~\onlinecite{supplementary}. From Fig.~\ref{training_1qubit} it follows that the 10-gate agent has demonstrated the best average reward during the training.

\subsection*{Experiments}
To examine the performance of the trained agent we performed experiments on  real quantum device provided by IBM. Fig.~\ref{efficiency} demonstrates these results on the level of the individual real quantum device experiments. One can see that the agent decreases the energy of the system starting from different random states and approaches to the $E$ = -0.75 defined as the average energy for the circuit (Fig.~\ref{ground-noise}b) simulating the exact solution of the problem. If the energy of the initial random state is low enough and close to the ground state the agent follows a passive strategy trying to keep such a winning position. On the other hand, a high-energy initial state is a signal to the agent to decrease the energy as much as possible from the currect position. More results and discussions can be found in Supplementary Material \cite{supplementary}.

\section*{Spin dimer}
\begin{figure}
	\includegraphics[width=\columnwidth]{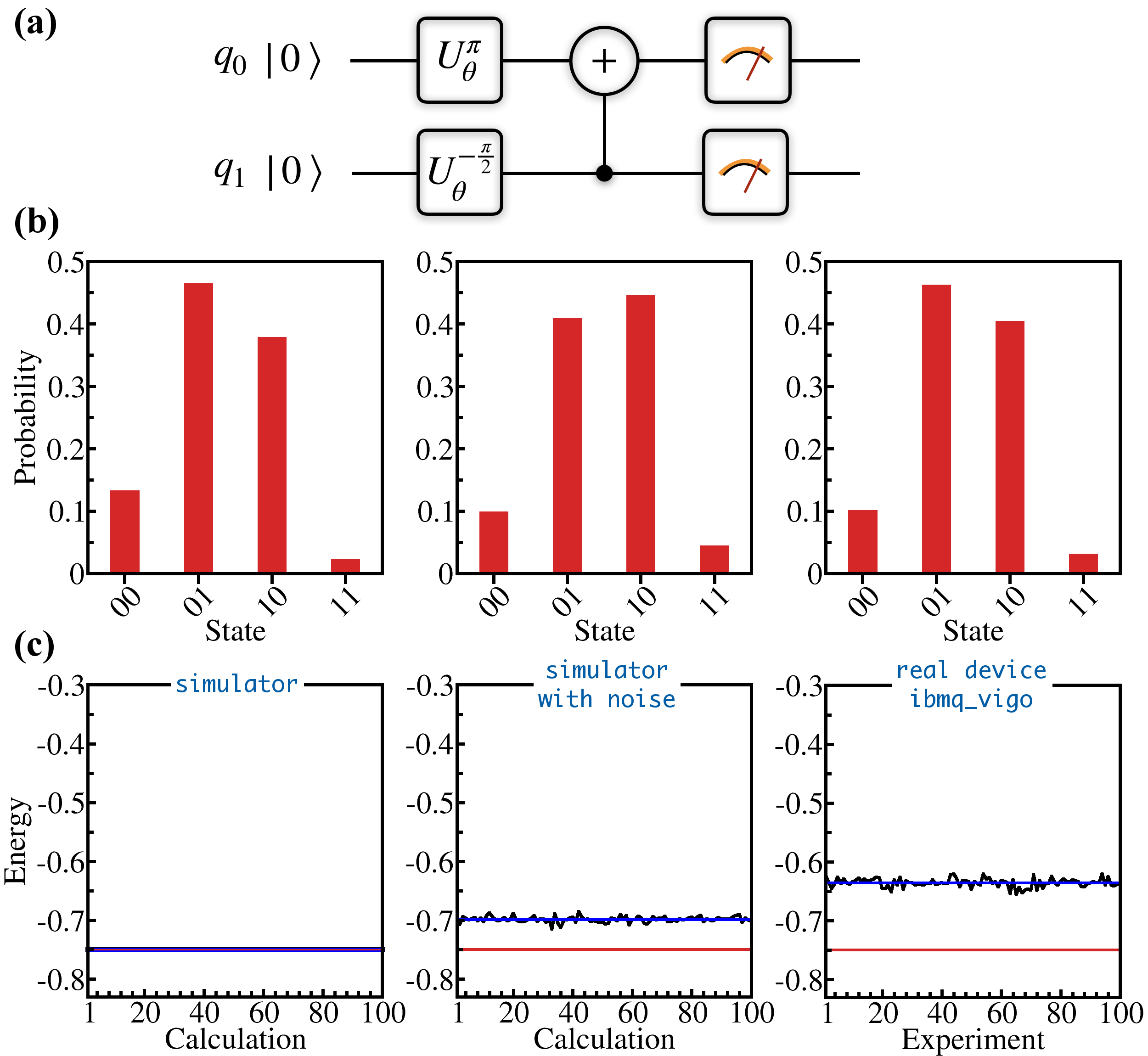}
	\caption{(a) Quantum circuit realizing singlet ground state for spin-$\frac{1}{2}$ dimer. Here $U_{\theta}^{\pi} = U_3(\pi, 0, 0)$ and $U_{\theta}^{-\frac{\pi}{2}} = U_3 (-\frac{\pi}{2}, 0, 0)$. (b) Probabilities of the basis states obtained from three independent experiments conducted on the real quantum device. (c) The energies obtained from 100 independent simulations on the quantum simulator (left and central panels) and 100 experiments on the real quantum device, IBM Q Vigo (right). The energy of exact solution denoted with red line is -0.75.}
	\label{dimer-noise}
\end{figure}

Having discussed the single-qubit problem we are in position to consider the agent's training in the two-qubit case. The corresponding spin model contains two antiferromagnetically-coupled spins, ${\rm \hat H} = J \hat {\bf S}_1 \hat {\bf S}_2 $. Here $J$ is isotropic exchange interaction. The ground state of this model is the singlet state $\ket{\Psi_0} = \frac{1}{\sqrt{2}} (\ket{\uparrow \downarrow} - \ket{\downarrow \uparrow})$. Such an entanglement state can be realized on a quantum computer with the circuit presented in Fig.~\ref{dimer-noise}\,(a). However, the experiments on the IBM Q Vigo (Fig.~\ref{dimer-noise}\,(b)) have shown significant contributions of the triplet states, $\ket{\uparrow \uparrow}$ and $\ket{\downarrow \downarrow}$. As a result the experimental ground state energy of -0.51 is considerably higher than the exact value of -0.75. It also explains disagreement between experiment and theory on ground state of the Heisenberg model defined on single square reported in Ref.~\onlinecite{VQE}. 

Another important observation is that an experiment conducted on the real IBM Q Vigo device gives the correlation functions that differ from the data of other experiments, which means that there are  fluctuations of the correlation functions and energies within the series of independent experiments (Fig.~\ref{dimer-noise}\,(c)).  Moreover, sets of the experiments conducted with the same circuits but at different periods of time can give different average energies \cite{supplementary}. 

\begin{figure*}
	\includegraphics[width=\linewidth]{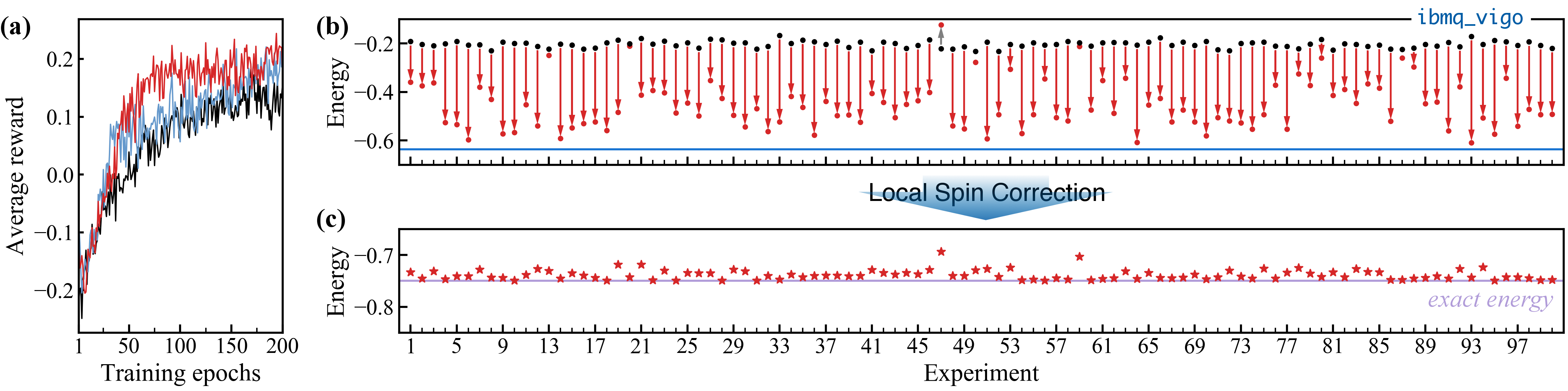}
	\caption{(a) The average reward achieved per circuit after the two-qubit agent is run with $\epsilon$-greedy policy ($\epsilon$=0.05) on the quantum simulator with noise. Black, red and blue curves denote training results obtained with elementary rotational angles of 0.5, 1.0 and 1.5 rad, respectively. (b) Performance of the trained agent demonstrated in experiments on the IBM Q Vigo device. Red arrows indicate lowering of energy with the corresponding circuits. Blue line shows the average energy (Fig.~\ref{dimer-noise}\,(c)) obtained with singlet state quantum circuit on the real device. The number of measurement shots was equal to 1024. (c) Ground state energies obtained on the real device after local moment correction. }
	\label{dimer-classic-eff}
\end{figure*}

In the two-qubit case the action list for the agent includes the same gates as for single-qubit agent and additional Controlled NOT (CNOT) gate that is responsible for the entanglement \cite{supplementary}.
The agent was trained starting with different random classical ground states. Such a non-entangled state is formed in the following way. The initial state of the first qubit is changed with random $U_3$ gate. Then the state of the second qubit is set to be antiparallel to the first one on the level of the Bloch sphere.  Interestingly, the agent has learned from self-play the possibility to overcome the classical energy limit by using the CNOT gate. A specific structure of the CNOT matrix implies an initial preparation by using single qubit rotations to decrease further the total energy with the corresponding entanglement gate. 

In Fig.~\ref{dimer-classic-eff}\,(a) we compare the rewards obtained within the training processes performed with different values of the elementary rotation angle, $\delta$ = 0.5, 1.0 and 1.5 rad (see Supplementary Material for further details on the rotation angle). One can see that the best and more stable results were obtained with $\delta$ = 1. In general the choice of the particular angle value can be also considered as a part of the reinforcement learning algorithm we propose. We left a practical realization of such an option for a future investigation.

Figure~\ref{dimer-classic-eff}\,(b) shows that the classical energy of the spin dimer estimated with quantum device is about -0.2, which is higher than the exact solution of -0.25.
The trained two-qubit agent decreases the energy of the system to the level of -0.6 (Fig.~\ref{dimer-classic-eff}\,(b)) that was obtained with the singlet state circuit. 

The average values denoted with blue lines in Fig.~\ref{efficiency} (Fig.~\ref{ground-noise}\,(c)) and Fig.~\ref{dimer-classic-eff}\,(b) (Fig.~\ref{dimer-noise}\,(c)) were obtained with the shortest quantum circuits (one-gate circuit for single-spin problem and three-gate circuit for dimer problem) imitating the exact solution. In the case of the reinforcement learning technique the agent constructs circuits of the 10-gate length. Thus, the neural network results obtained with longer circuits closely approach to the level of the short circuits data, which is an additional demonstration of the performance of our method.

\begin{figure}
	\includegraphics[width=\columnwidth]{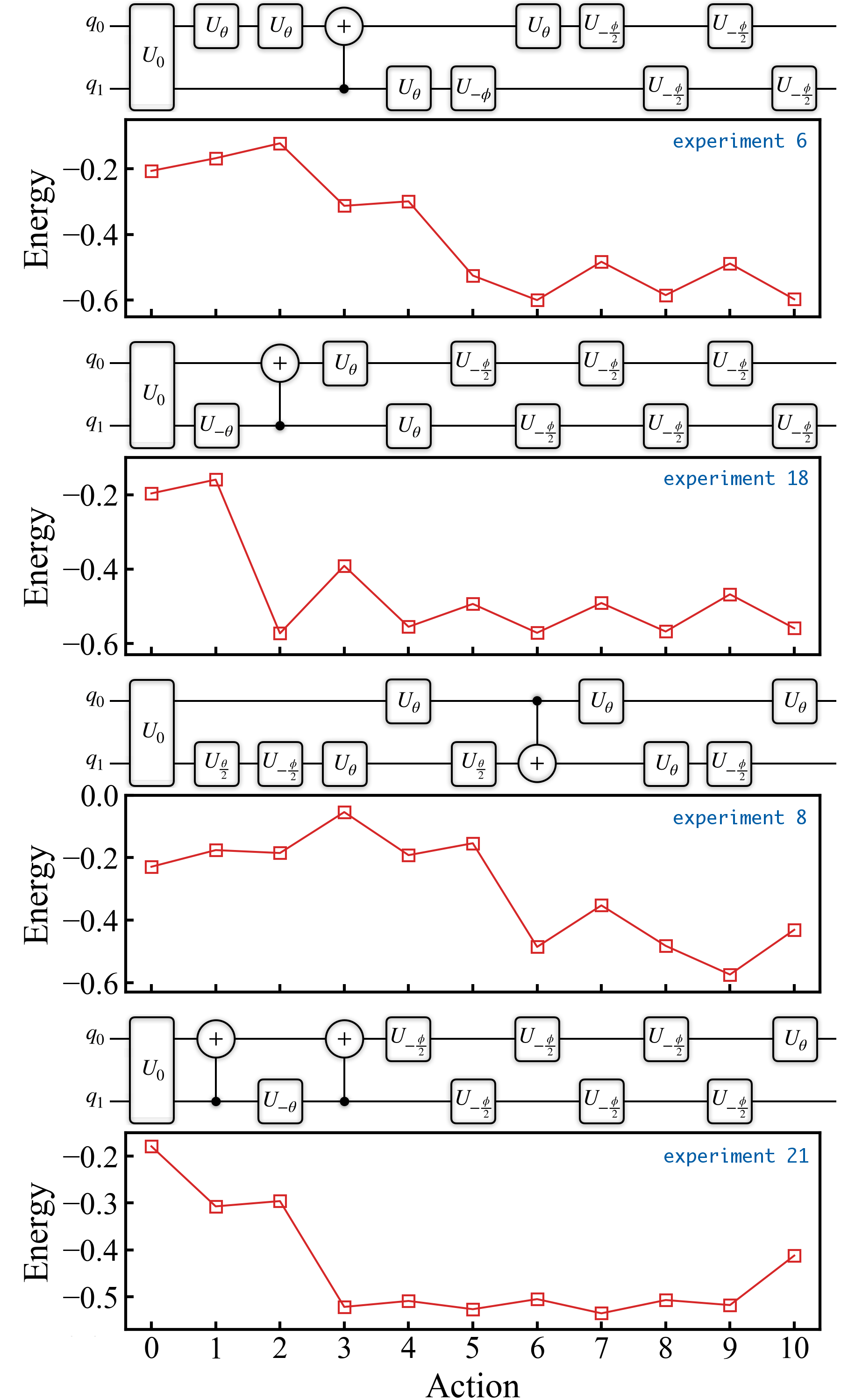}
	\caption{Examples of quantum system energy evolution during some experiments presented in Fig.~\ref{dimer-classic-eff}. These 10-gate circuits were constructed gate-by-gate by the neural network agent trained to approximate the spin Hamiltonian ground state.}
	\label{circuit}
\end{figure}

It is also important to analyze the evolution of the energy obtained with neural network during individual experiments. Some examples given in Fig.~\ref{circuit} demonstrate that the largest energy decreases are mainly achieved with the CNOT gate which can be implemented not only in the beginning of the circuit construction.
To keep a minimal energy the agent uses the rotation gates from the action list with smallest angles, such as $\frac{\phi}{2}$. It can be seen by the examples of the experiments 6, 18 and 21.

\subsection*{Local spin correction}
As it was shown above, the direct imitation of the single ground state as well as its neural network approximation on the real quantum device are characterized by the energies which are considerably larger than the exact ground state energy. Such a disagreement between experiment and theory is mainly related to   decoherence and gate errors. In this situation it is important to find a way to compensate such hardware imperfections. The different strategies can be used for that. The authors of Ref.~\onlinecite{Carretta} have simulated time-dependent spin-spin correlation functions of the Heisenberg-type magnetic systems in high magnetic fields. At these magnetic fields the quantum ground state of the spin system is fully polarized. To fit the experimental results obtained with an IBM quantum computer to exact ones the authors introduced a phase-and-scale procedure that is based on an artificial phase correction of the local spin-spin correlation function at the zero time and application of the same correction to the whole time domain.  Below we show that for the quantum systems in the singlet ground state a similar correction procedure can be derived from a general sum rule for the spin-spin correlation functions.    

For the quantum systems characterized by a singlet-type ground state we propose a local moment correction procedure that systematically improves the agreement between experiment and theory on the ground state energy. The ground state spin-spin correlation functions of a quantum spin system with even number of spins and antiferromagnetic interaction between them satisfies to the following sum rule 
\begin{eqnarray}
\sum_{ij} \langle \hat {\bf S}_i \hat {\bf S}_j \rangle = 0. 
\end{eqnarray}
The sum in this equation contains local and non-local contributions that can be decomposed. It gives
\begin{eqnarray}
\sum_{i} \langle (\hat {\bf S}_i)^2 \rangle = -\sum_{i \ne j} \langle \hat {\bf S}_i \hat {\bf S}_j \rangle.
\label{local_nonlocal} 
\end{eqnarray}
The sum in the right part contains all possible pair correlation functions in the system. Namely these non-local correlation functions are calculated at each iteration of our algorithm with quantum computer and used as input for neural network. Within the correction procedure we use them to estimate the local  correlator in the left part of Eq.~\eqref{local_nonlocal} that should be compared with the exact value $S(S+1)$. Namely, this exact value is important insider information for us on unknown quantum system. The ratio 
\begin{eqnarray}
X = \frac{S(S+1)}{\frac{1}{N} \sum_{i} \langle  (\hat {\bf S}_i)^2 \rangle}
\label{correction_multiplier}
\end{eqnarray}
can be used to correct the non-local correlation functions $ X \langle \hat {\bf S}_i \hat {\bf S}_j \rangle$ ($N$ is the number of spins). It gives opportunity to correct the estimated energy of the Heisenberg system. In the case of the dimer system we have a trivial situation since the nonlocal spin-spin correlation functions $ \braket{\hat{\mathbf{S}}_1\hat{\mathbf{S}}_2} $ and $ \braket{\hat{\mathbf{S}}_2\hat{\mathbf{S}}_1} $ fully define the on-site spin-spin correlators $ \braket{\hat{\mathbf{S}}_1^2} $ and $ \braket{\hat{\mathbf{S}}_2^2} $. The modified experimental results are presented in Fig.~\ref{dimer-classic-eff}\,(c). One can see that the most part of the experiments gave very accurate estimation of the exact energy of the spin dimer. 

The procedure we propose is of general nature and can be used not only in the case of the neural network solver presented in this work but also in conjunction with other quantum computer eigensolvers.

\subsection*{Comparison with variational quantum eigensolver}
Since we propose a new neural network eigensolver it is important to implement previously developed quantum computer approaches to the same ground state problem of the Heisenberg dimer and compare the performance of these methods. As we have discussed in the introduction there are two standard quantum computer eigensolvers, they are variational quantum algorithm and phase estimation method. The comparison of these methods can be found in Ref.~\onlinecite{Malley} in which the authors performed quantum computer experiments and computed the energy of hydrogen molecule. For the quantum computer experiments this electronic structure problem is formulated in a form of a 2-spin Hamiltonian similar to that we consider in this work.  It was demonstrated that the variational quantum approach outperforms the phase estimation algorithm. That is why we perform the comparison the former method with our neural network solver by the example of the spin dimer model. For that a realization of the variational approach reported in Ref.~\onlinecite{VQE} was used. Within this approach the wave function of a quantum Heisenberg model is stored on a quantum device and represented in the following form
\begin{eqnarray}
|\Psi ({\boldsymbol \theta}) \rangle = U_{\rm CNOT} U_{3}({\boldsymbol \theta}) |00 \rangle,
\end{eqnarray}
where $|00 \rangle$ is the initial state of the 2-qubit system, $U_{3}$ is a set of rotation gates, $U_{\rm CNOT}$ is the CNOT gate that is responsible for the entanglement and $\theta$ is the set of angles we vary to approximate the ground state of the spin Hamiltonian. The angles at the $(k+1)$th iteration are defined as 
\begin{eqnarray}
{\boldsymbol \theta}_{k+1} = {\boldsymbol \theta}_{k} - \alpha_k {\bf g}_k ({\boldsymbol \theta}_k),
\label{var_theta}
\end{eqnarray}
here ${\bf g}_k ({\boldsymbol \theta}_k)$ is the gradient at ${\boldsymbol \theta}_k$ and $\alpha_k$ is the step size parameter. This variational scheme contains a number of parameters for which we use the same values as it was proposed in Ref.~\onlinecite{VQE}. Every 20 steps we perform a calibration procedure to probe the energy landscape at the quantum state defined by the current set of the rotation angles and to renew the gradient-descent parameters.   

\begin{figure}
	\includegraphics[width=\columnwidth]{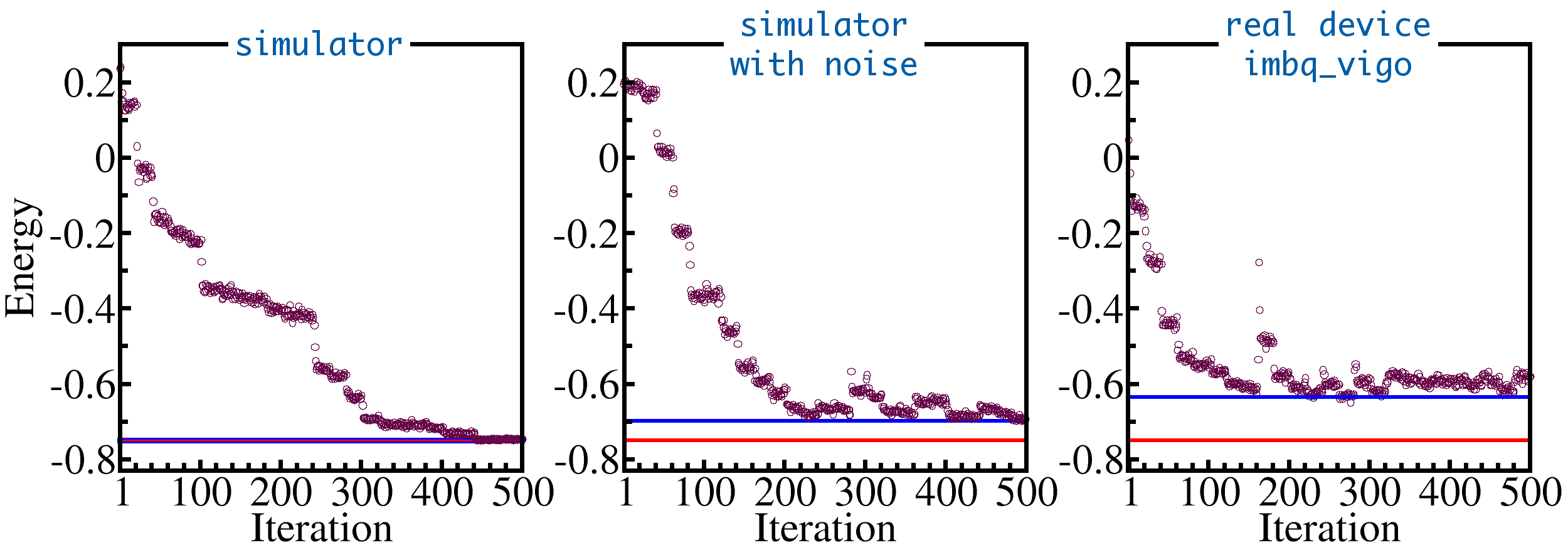}
	\caption{Results of calculations and experiments obtained variational quantum eigensolver realized on the basis of work \cite{VQE}.}
	\label{variational}
\end{figure}

The results obtained by using the variational approach with simulator, simulator with noise and real quantum device are presented in Fig.~\ref{variational}. In the case of the ideal quantum simulator we observe excellent agreement between the calculated energies and exact results. The average energy estimated with noisy simulator is -0.7, which is larger than the exact solution of -0.75. In turn, the real quantum device IBM Q Vigo gives the energy of -0.6. It agrees with neural network results (Fig.~\ref{dimer-classic-eff}\,(b)). The real experiments are characterized by fluctuations of the energy after 200 iterations. They are due to large values of the step size $\alpha$ in Eq.~\eqref{var_theta} obtained with calibration procedure. Here, one of the possible solutions is to limit the upper bound of $\alpha$ to 2. 

Despite the agreement on real device results of variational and neural network solvers there are important differences between them. First of all the neural network results were obtained with circuits of 10 gates. At the same time there are 3 gates (two universal $U_3$ gates and one CNOT gate) in the variational scheme. Taking into account that gate errors accumulate as circuit length increases, our neural network approach seems to be stable to such errors. Another important difference is the initial quantum state. For neural network solver we always start with a random initial quantum state, which suggest the way to avoid local energy minima for more complicated spin Hamiltonian problems.  

\section*{Conclusion}
We have developed a neural-network agent approximating the ground state of a spin model on a quantum device. The agent was trained by reinforcement learning from self-play on quantum simulator with noise. Here the general objective of our reinforcement learning approach is to obtain the agent that can play moderately well in all the cases even on noisy real quantum devices. The agent performance was demonstrated on quantum simulator and real quantum devices. In the case of the dimer problem we found that the agent can learn entanglement by applying the CNOT gate. In combination with local spin correction our neural network approach provides excellent agreement with exact dimer solution on the ground state energy.

To consider the systems with more than two qubits, a new strategy should be adopted. For instance, one can use a single deep neural network for all the gates from the action list. In the current formulation of the neural network approach the set of observables will increase exponentially with the system size. Moreover, to compensate the decoherence we need to define all the nonlocal spin-spin correlation functions to get an accurate estimation of the ground state energy. This is the problem if we use classical computers to simulate the neural network.  
The problem could be solved if the network is realized with another quantum device. Recent studies \cite{quantum_network1, quantum_network2} have demonstrated that the equivalent of m-dimensional classical input and weight vectors can be encoded on the quantum hardware by using N qubits, where m = $2^N$. Similar approach allows to exploit the exponential advantage of quantum information storage. 

Another important problem is a smart selection of the implemented gates, since their number substantially grows as the system size increases. Such a selection can be also fulfilled with machine learning approach \cite{PNAS}.

\begin{acknowledgments}
	This work was supported by the Russian Science Foundation, Grant No. 18-12-00185.
\end{acknowledgments}


\end{document}


\begin{center}
		\textbf{\large Supplemental Material: Neural network agent playing spin Hamiltonian games on a quantum computer}
	\end{center}
	\setcounter{equation}{0}
	\setcounter{figure}{0}
	\setcounter{table}{0}
	\setcounter{page}{1}
	\makeatletter
	\renewcommand{\theequation}{S\arabic{equation}}
	\renewcommand{\thefigure}{S\arabic{figure}}

	\section*{Training algorithm}
	In this section we present technical details concerning the reinforcement learning approach. The basic steps of the algorithm we use is given in the listing below
	\begin{lstlisting}
Initialize replay memory ?$D$? with capacity M
Initialize action-value function ?$Q$? with random weights ?$\xi_0 (a_k)$? for each action gate ?$a_k$?
Initialize target action-value function ?$\hat Q$? with weights ?$\xi^{-}(a_k) = \xi_0(a_k)$?
?{\bf For}? circuit = 1, Num_circuits ?{\bf do}? 
  Add random ?$U_3$? gate, perform the measurement and initialize ?$\text{\normalsize\rmfamily\textbf{s}}_1$?
  ?{\bf For}? t = 1, Num_gates ?{\bf do}?
    With probability of ?$\epsilon$? select a random gate ?$a_{t}$?
    otherwise select ?$a_t = {\rm argmax}_{a} Q(\text{\normalsize\rmfamily\textbf{s}}_{t},a;\xi (a))$?
    Add gate ?$a_t$? to the circuit, perform the measurement, form ?$\text{\normalsize\rmfamily\textbf{s}}_{t+1}$? and estimate the reward ?$r_t$?
    Store the transition (?$\text{\normalsize\rmfamily\textbf{s}}_{t}, a_{t}, \text{\normalsize\rmfamily\textbf{s}}_{t+1}, r_{t}$?) in ?$D$?
    Sample random minibatch storing a transition (?$\text{\normalsize\rmfamily\textbf{s}}_{\tau}, a_{\tau}, \text{\normalsize\rmfamily\textbf{s}}_{\tau+1}, r_{\tau}$?) from ?$D$?
    ?\begin{equation*}
      y_{\tau} =  \left\{  \begin{aligned}
      &r_\tau \quad \text{if circuit terminates at step $\tau$+1 (Num\_gates=$\tau$+1)} \\
      &r_\tau + \gamma {\rm max}_{a'} \hat Q(\text{\normalsize\rmfamily\textbf{s}}_{\tau+1}, a';\xi^{-} (a')) \quad \text{otherwise}
      \end{aligned} \right. 
    \end{equation*}?
    Perform gradient descent step on ?$(y_\tau - Q(\text{\normalsize\rmfamily\textbf{s}}_\tau, a_\tau; \xi_{t}(a_{\tau})))^2$? with respect to ?$\xi_{t}(a_\tau)$?
    Every C steps reset ?$\xi^{-}(a_k) = \xi_t(a_k)$?
  ?{\bf End}? for
?{\bf End}? for
\end{lstlisting}
	
	The parameters used for training the neural networks are described in Table \ref{parameters}
	
	\begin{table}[!h]
		\centering
		\caption{Parameters of the reinforcement learning used to train the agent.\label{parameters}}
		\begin{tabular}{ccc}
			\hline
			Parameter & Value  & Description  \\
			\hline
			replay memory size, $M$ & 32 & Number of minibatches storing the transitions  ($\mathbf{s}_{t}, a_{t}, \mathbf{s}_{t+1}, r_{t}$) \\
			global learning rate, $\alpha$ & 0.05 & the learning rate used in \eqref{wi} \\
			target network update frequency, $ C $ & 500 &  Number of parameters updates with which target state is updated \\
			number of hidden neurons & 32 (64) & Neuron number for one-qubit (two-qubit) problems \\
			discount factor, $\gamma$ & 0.99 & discount factor in \eqref{lossfunction} \\
			Num\_circuits & 100 & Number of the constructed circuits for one epoch \\
			Num\_gates & 5, 10, 15 & Maximal number of gates in one circuit \\
			initial exploration & 1 & Initial value of $\epsilon$ in $\epsilon$-greedy exploration \\
			final exploration & 0.05 & Final value of $\epsilon$ in $\epsilon$-greedy exploration \\
			final exploration step & 10 $\times$ Num\_gates & \makecell{Number of measurements over which the initial value of $\epsilon$\\is annealed to its final value} \\
			\hline
		\end{tabular}
	\end{table}
	
	The $\epsilon$-greedy policy described in Ref.~\onlinecite{Sgame} was used in our study. With probability $\epsilon$ the agent selects a random gate and with probability 1-$\epsilon$ it follows greedy policy $a_t = {\rm argmax}_{a} Q({\bf s}_{t},a; \xi_t (a))$. The latter expression means that for specific system state ${\bf s}_{t}$ the agent will choose a gate for which the neural network gives the largest action value function Q. 
	
	Following the work \onlinecite{Sgame} we introduce two sets of weights for each action, $\xi$ and $\xi^{-}$ at each iteration. Thus, for specific action $a_{i}$ and specific iteration $t$ we use the notation $\xi_t (a_k) = \{W_{ij}^h(a_k), W_i^o(a_k)\}_t$. The loss function at $t$th iteration is given by the following expression
	\begin{eqnarray}\label{lossfunction}
	L_t (\xi_t (a_k)) = [r + \gamma {\rm max}_{a'} \hat Q({\bf s}_{\tau+1}, a';\xi^{-} (a')) - Q ({\bf s}_{\tau}, a_k, \xi_t (a_k))]^2
	\end{eqnarray}
	
	Since the values of the output neurons are in the range from -1 to 1, we clip $y_{\tau}$ to be between -1 and 1.
	
	\section*{Neural network training}
	As an input for neural networks we use the correlation functions calculated with quantum simulator with noise emulation. The input and output of the hidden layer neurons are given by
	
	\begin{equation}
	h^{out}_j=\frac{1}{1+e^{-h^{inp}_j}},\quad h^{inp}_j= \sum_{i=1}^{N_{inp}}s_i W^h_{ij},
	\end{equation}
	where $s_i$ is the value of $i$th input neuron ($i$th correlation function), $W^h_{ij}$ --- weight between the $i$th input neuron and $j$th hidden neuron, $N_{inp}$ --- number of the input neurons which is equal to number of the correlation functions calculated with quantum computer. The value of the output neuron is calculated in a standard way by using the following equations
	
	\begin{figure}[t] 
		\centering
		\includegraphics[width=0.4\columnwidth]{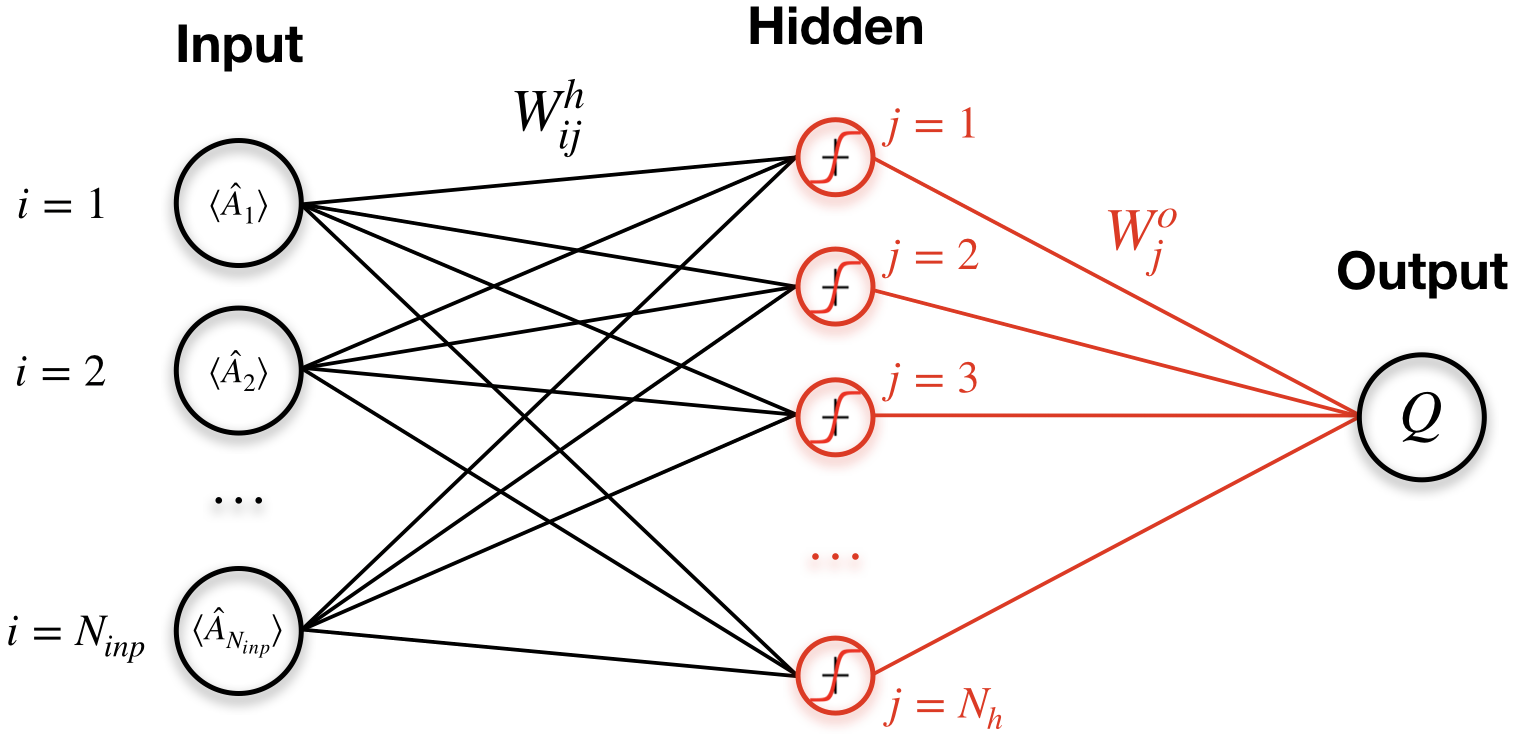} 
		\caption{Schematic representation of constructed neural network with single hidden layer. We used sigmoid as an activation function of hidden and output neurons. All the notations are described in the text below.}
		\label{netw} 
	\end{figure}
	
	\begin{equation}\label{oinp}
	o^{out} = 2 \left(\frac{1}{1+ e^{-o^{inp}}}-0.5\right),\quad o^{inp}= \sum_{j=1}^{N_h}h^{out}_j W^o_{j},
	\end{equation}
	where $N_h$ is the number of hidden neurons, $W^o_{j}$ --- weight between the $j$th hidden neuron and output neuron. Here we use symmetric version of the sigmoid function.
	
	The error function is given by
	\begin{equation}\label{Error}
	E[\xi_t (a_k)] = (o^{target}-o^{actual}[\xi_t (a_k)])^2,
	\end{equation}
	where $o^{target}$ and $o^{actual}$ are the target and actual output value of the tailing neuron, \eqref{oinp}.
	
	Due to the fact that we optimized our network through back-propagation method by means of the gradient descent we used the following expression for new weights 
	\begin{equation}\label{wi}
	W (t+1)= W(t) + \alpha \cdot g \cdot \nabla_{\xi_t}E,
	\end{equation}
	where $\alpha$ and $g$ are the global and individual learning rates, respectively. With the gain parameter $g$ we control the individual learning rates for each weight of each network. The gain are constrained to lie in the range [0.1, 2]. The local gain is increased, $g(t) = g(t-1) + 0.05$ if the gradient for that weight does not change the sign, $\nabla_{\xi_t}E(t)\nabla_{\xi_t}E(t-1) > 0$. Otherwise, $g(t) = g(t-1) \cdot 0.95$.  
	
	The weights gradients were calculated by the following equations 
	\begin{equation}\label{dwo}
	\frac{\partial E}{\partial W_j^o} = \delta o \cdot h^{out}_j
	\end{equation}
	and
	\begin{equation}\label{dwh}
	\frac{\partial E}{\partial W_{ij}^h} = \delta h^{out}_j \cdot s_i,
	\end{equation}
	where
	\begin{equation}\label{do}
	\delta o= (o^{target}-o)\cdot (o+1)\cdot (\frac{1}{2}-\frac{o}{2}),
	\end{equation}
	\begin{equation}\label{dh}
	\delta h^{out}_j= h^{out}_j(1-h^{out}_j)W^o_{j}\delta o.
	\end{equation}
	
	\section*{Calculation of the energy}
	To define the energy of the system in question we use the correlation functions that can be calculated through the  probabilities of the different basis states, $P$. The latter is a standard output of a quantum computer. In a one-qubit case the correlation function $\braket{Z} = P(0) - P(1)$, where $P(0)$ and $P(1)$ are the probabilities of the state $0$ and $1$, respectively. To measure the $\braket{X}$ and $\braket{Y}$ one is to rotate the standard basis frame to lie along the corresponding axis, and to make a measurement in the standard basis \cite{SIBMQE}. For instance, for the $\braket{X}$ correlator we implement Hadamard (Fig.~1, main text) gate before the measurement gate.
	
	In the case of the dimer problem we are interested in the $\braket{Z_1Z_2}$, $\braket{X_1 X_2}$, $\braket{Y_1 Y_2}$, $\braket{X_1 Y_2}$, $\braket{Y_1 X_2}$, $\braket{X_1 Z_2}$, $\braket{Z_1 X_2}$, $\braket{Y_1 Z_2}$, $\braket{Z_1 Y_2}$ correlation functions, since we use them as input for our neural network agent. The $zz$ correlator is to be defined by using the following combination of the basis states probabilities:
	$\braket{Z_1 Z_2} = P(00) + P(11) - P(01) - P(10)$. To determine other correlation functions one needs to perform the rotations of the standard basis frame as described above.
	
	\section*{Comparison of the simulator with noise and real quantum device}
	
\begin{figure}[!h]
		\centering
		\includegraphics[width=0.7\columnwidth]{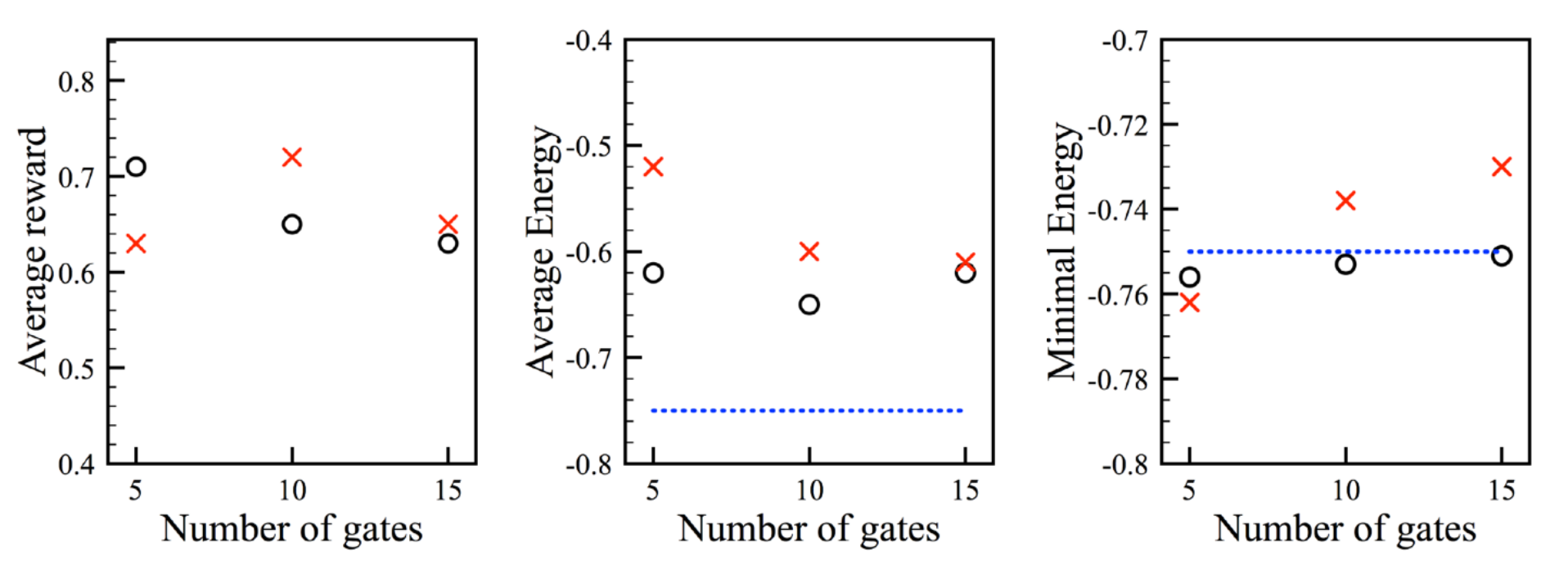}
		\caption{Performance of the trained one-qubit agent estimated with noisy quantum simulator (black circles) and real quantum device (red crosses). Left, center and right panels represent the average reward, average energy and minimal energy obtained with quantum circuits of 5, 10 and 15 gates. The dashed blue line denotes the average energy obtained with circuits corresponding to the exact solution of the single spin problem.}
		\label{result-1qubit}
	\end{figure}

	To examine the performance of the trained one-qubit agent we performed experiments on quantum simulator and real quantum device provided by IBM Quantum Experience. During these experiments the agent was approximating ground state of the single-spin Hamiltonian. Here we used the neural networks weights obtained for 10-gate agent demonstrating the best average reward. The results obtained with 5, 10 and 15 gates quantum circuits are presented in Fig.~\ref{result-1qubit}. In analogy with the Google's agent playing classical Atari games, we detect the best score that is the minimal energy in our case (Fig.~\ref{result-1qubit} right panel). Remarkably, in all the cases the energies obtained by the agent on the real device and simulator are close to that estimated with quantum circuit simulating the exact solution. It confirms importance of the noise model we used for realistic simulation of the experimental data.
	
	\section*{Stability of the quantum computer measurements over time}
\begin{figure}[!h]
		\centering
		\includegraphics[width=0.82\linewidth]{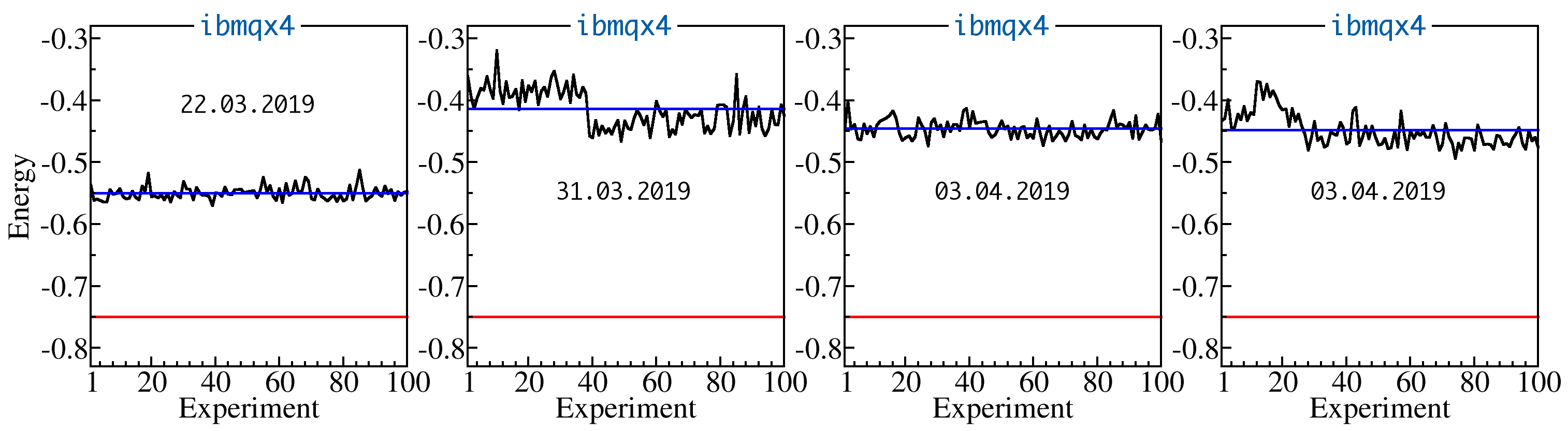}
		\caption{Energies obtained from the real device experiments conducted on different days. In all the experiments the same quantum circuit simulating singlet state (Fig.~6, main text) was used. The blue lines denote the average energy for each set of the experiments. The red line corresponds to the exact ground state energy for the spin dimer.}
		\label{dimer-exat-noise}
	\end{figure}

	Here, by the example of the dimer problem we give an additional demonstration that the agent explores a very complex and dynamical computational environment. The results of the real experiments conducted with singlet state circuit (Fig.~6, main text) are presented in Fig.~\ref{dimer-exat-noise}. One can see that the average energies estimated on different days vary from -0.55 to -0.4.
	
	\section*{Gate description}
	Assuming that the state of $ n $th qubit is written in form:
	\begin{equation}
	\ket{\phi_n} = \cos\frac{\theta_n}{2}\ket{0} + \cos\frac{\theta_n}{2}e^{-i\phi_n}\ket{1}
	\end{equation}
	one can represent the generic operator $ U_3 $ as the following:
	\begin{equation}
	U_3(\theta, \phi, \lambda) = 
	\begin{pmatrix}
	\cos\frac{\theta}{2} & e^{-i\lambda}\sin\frac{\theta}{2} \\
	e^{i\phi}\sin\frac{\theta}{2} & e^{i\lambda + i\phi}\cos\frac{\theta}{2}\\
	\end{pmatrix}
	\end{equation}
	
	The gates that can be used by the one- and two-qubit agents are listed in Table \ref{gates}. We have found that the optimal value of $ \delta $ that yields the best reward during calculation is equal to 0.5 and 1 rad for single spin and dimer problems respectively.
	
	\begin{table}[!t]
		\centering
		\caption{List of one- and two-qubit gates used in this work.}\label{gates}
		\begin{tabular}{|c|c|c||c|c|c|}
			\hline
			Action & Gate & Description & Action & Gate & Description \\\hline
			\begin{minipage}{0.05\columnwidth}\includegraphics[width=\linewidth]{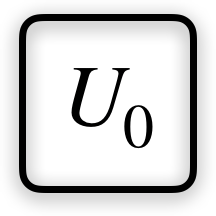}\end{minipage} & $\ U_3(\delta^{rand}, \phi^{rand}, \lambda^{rand})\ $ & \makecell{rotation\\by random angle} &
			
			\multirow{2}{*}{\begin{minipage}{0.05\columnwidth}\includegraphics[width=\linewidth]{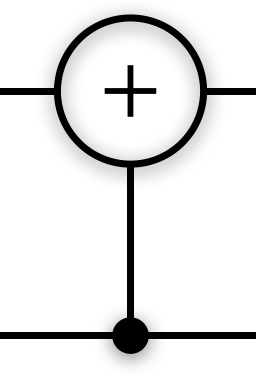}\end{minipage}} & \multirow{2}{*}{\makecell{\\[5px]$ CNOT $}} & \multirow{2}{*}{\makecell{\\exclusive NOT\\(creation of entanglement)}} \\
			
			\begin{minipage}{0.05\columnwidth}\includegraphics[width=\linewidth]{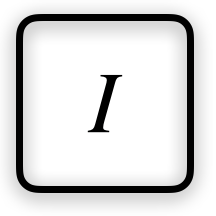}\end{minipage} & $ IDLE $ & do nothing &  & &  \\
			
			\begin{minipage}{0.05\columnwidth}\includegraphics[width=\linewidth]{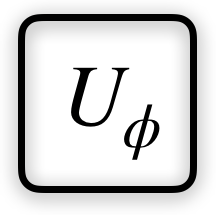}\end{minipage} & $ U_3(0,\delta,0) $ & \makecell{rotation\\around Z axis by $ \delta $} & 
			
			\begin{minipage}{0.05\columnwidth}\includegraphics[width=\linewidth]{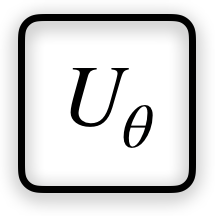}\end{minipage} & $ U_3(\delta,0,0) $ & \makecell{rotation\\around Y axis by $ \delta $} \\
			
			\begin{minipage}{0.05\columnwidth}\includegraphics[width=\linewidth]{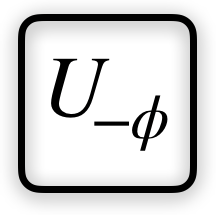}\end{minipage} & $ U_3(0,-\delta,0) $ & \makecell{rotation\\around Z axis by $ -\delta $} & 
			
			\begin{minipage}{0.05\columnwidth}\includegraphics[width=\linewidth]{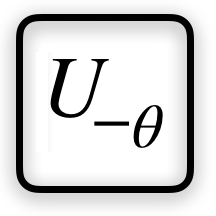}\end{minipage} & $ U_3(-\delta,0,0) $ & \makecell{rotation\\around Y axis by $ -\delta $} \\
			
			\begin{minipage}{0.05\columnwidth}\includegraphics[width=\linewidth]{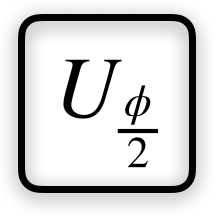}\end{minipage} & $ U_3(0,\frac{\delta}{2},0) $ & \makecell{rotation\\around Z axis by $ \frac{\delta}{2} $} &
			
			\begin{minipage}{0.05\columnwidth}\includegraphics[width=\linewidth]{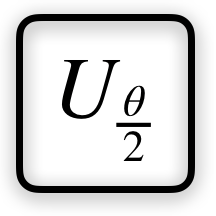}\end{minipage} & $ U_3(\frac{\delta}{2},0,0) $ & \makecell{rotation\\around Y axis by $ \frac{\delta}{2} $} \\
			
			\begin{minipage}{0.05\columnwidth}\includegraphics[width=\linewidth]{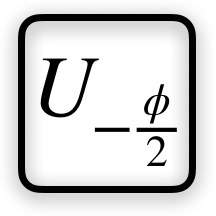}\end{minipage} & $ U_3(0,-\frac{\delta}{2},0) $ & \makecell{rotation\\around Z axis by $ -\frac{\delta}{2} $} &
			
			\begin{minipage}{0.05\columnwidth}\includegraphics[width=\linewidth]{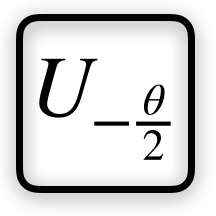}\end{minipage} & $\ U_3(-\frac{\delta}{2}, 0,0)\ $ & \makecell{rotation\\around Y axis by $ -\frac{\delta}{2} $} \\
			\hline\hline
		\end{tabular}
	\end{table}